\documentclass[a4paper,11pt,fleqn]{article}

\usepackage[totalwidth=465pt,totalheight=645pt]{geometry}

\usepackage[latin1]{inputenc}
\usepackage{amsmath,amssymb,amsthm,bbm,graphicx,rcs,fancyhdr}
\usepackage{natbib}
\bibpunct{(}{)}{,}{a}{,}{,}
\RCS $Id: article.tex,v 1.36 2011-02-14 21:24:44 cappe Exp $

\newcommand{\1}{{\mathbbm{1}}}
\newcommand{\E}{{\operatorname{E}}}
\renewcommand{\P}{{\operatorname{P}}}

\newcommand{\ut}{{\mathrm{t}}}

\newcommand{\eqsp}{\medspace}
\newtheorem{theorem}{Theorem}
\newtheorem{proposition}{Proposition}
\newtheorem{lemma}{Lemma}
\newtheorem{corollary}{Corollary}
\newtheorem{algorithm}{Algorithm}
\newtheorem{assumption}{Assumption}
\theoremstyle{remark}

\title{Online EM Algorithm for Hidden Markov Models}
\author{Olivier Capp\'{e}}
\date{LTCI, Telecom ParisTech \& CNRS}

\begin{document}

\maketitle

\begin{abstract}
  Online (also called ``recursive'' or ``adaptive'') estimation of fixed model
  parameters in hidden Markov models is a topic of much interest in times
  series modelling. In this work, we propose an online parameter estimation
  algorithm that combines two key ideas. The first one, which is deeply rooted
  in the Expectation-Maximization (EM) methodology consists in reparameterizing
  the problem using complete-data sufficient statistics. The second ingredient
  consists in exploiting a purely recursive form of smoothing in HMMs based on
  an auxiliary recursion. Although the proposed online EM algorithm resembles a
  classical stochastic approximation (or Robbins-Monro) algorithm, it is
  sufficiently different to resist conventional analysis of convergence. We
  thus provide limited results which identify the potential limiting points of
  the recursion as well as the large-sample behavior of the quantities involved
  in the algorithm. The performance of the proposed algorithm is numerically
  evaluated through simulations in the case of a noisily observed Markov
  chain. In this case, the algorithm reaches estimation results that are
  comparable to that of the maximum likelihood estimator for large sample
  sizes.

\

\textbf{Keywords} Hidden Markov Models, Expectation-Maximization Algorithm, Online Estimation, Recursive Estimation, Stochastic Approximation, Smoothing
\end{abstract}


\section{Introduction}
Hidden Markov modelling is a key concept of statistical time series analysis, which has had a wide-ranging practical impact over the latest forty years.
Hidden Markov models (HMMs) in their classical
form (i.e., when the state variable is finite-valued) are sufficiently simple to give
rise to efficient inference procedures while allowing for useful modelling of
various practical situations. Ever since the pioneering contributions of
\cite{baum:eagon:1967,baum:petrie:soules:weiss:1970}, the EM
(Expectation-Maximization) algorithm has been the method of choice for
parameter inference in HMMs. The EM algorithm is a dedicated numerical
optimization routine which aims at maximizing the (log) likelihood of a batch
of observations. It tends to be preferred to its alternatives due to its
robustness and ease of implementation.

This contribution is devoted to online parameter estimation for HMMs, in which the available
observations are only scanned once and never stored, allowing for a continuous
adaptation of the parameters along a potentially infinite data stream. In the
case of HMMs, online parameter estimation is a challenging task due to the
non-trivial dependence between the observations. The EM-inspired methods proposed so far have been either based on
finite-memory approximations of the required smoothing computations
\citep{krishnamurthy:moore:1993} or on finite-memory approximations of the
data log-likelihood itself \citep{ryden:1997}. An alternative consists in
using gradient-based methods \citep{legland:mevel:1997} which do not directly
follow the principles of the EM algorithm. \cite{kantas_doucet_singh_maciejowski_09} provide a comprehensive recent review of these methods, including more advanced aspects for models that require the use of simulation-based methods.
Recently, \cite{mongillo:deneve:2008}
proposed an online version of the EM algorithm for HMMs in the case where both
the states and observations take a finite number of values. The key ingredient
of this algorithm is a recursion which allows for recursive computation of
smoothing functionals required by the EM algorithm. However, this recursion
appears to be very specific and its potential application to more general types
of HMMs is not addressed by \cite{mongillo:deneve:2008}.

The purpose of this paper is to build on the idea of
\cite{mongillo:deneve:2008} in light of the framework introduced
by~\cite{cappe:moulines:2009} for online EM estimation in the case of
independent observations. The first contribution of the paper is an algorithm
that extends the proposal of \cite{mongillo:deneve:2008} to general HMMs, with
possibly continuous observations. This algorithm is based on the key
observation that the recursion used in \cite{mongillo:deneve:2008} is an
instance of the recursive smoothing scheme for sum functionals introduced
by~\cite{zeitouni:dembo:1988,elliott:aggoun:moore:1995}. Although a complete
analysis of the proposed algorithm is currently lacking, we provide a first
result that identifies the possible limiting points of the algorithm. These
coincide with the stationary points of a limiting EM mapping that may be
interpreted as the limit of the EM recursion under an infinite number of
observations. This interpretation which generalizes the argument
of~\cite{cappe:moulines:2009} for the case of independent observations also
provides some interesting insight regarding the behavior of the batch EM
algorithm when used with a large number of observations.

The remaining of the paper is organized as follows. Section~\ref{sec:alg:HMMs}
opens with a brief review of our modelling assumptions and of smoothing
computations in HMMs. The proposed online EM algorithm is then introduced in
Section~\ref{sec:online}. Section~\ref{sec:discuss} is devoted to a discussion
of the online EM algorithm and, in particular, of its connections with previous
works and of its numerical complexity. Section \ref{sec:results} contains
preliminary results pertaining to the convergence of the method, with
corresponding proofs to be found in the appendix. Finally, in Section
\ref{sec:gausHMM} we apply the online EM algorithm to the estimation of the
parameters of a Markov chain observed in Gaussian noise and illustrate its
performance through numerical simulations.

\section{Online EM Algorithm for HMMs}
\label{sec:alg:HMMs}

\subsection{Model and Notations}
It is assumed that the state and observation sequences,
$(X_t,Y_t)_{t\in\mathbb{Z}}$ are generated under a \emph{stationary} Hidden
Markov model with unknown parameter $\theta_\star$, where $(X_t)$ takes its
values in some finite set $\mathcal{X}$. The notations $\ell_{\theta_\star}$,
$p_{\theta_\star}$, $\P_{\theta_\star}$ and $\E_{\theta_\star}$ refer to,
respectively, the likelihood, the joint density of the states and observations,
the probability, and, the expectation under the model parameterized by
$\theta_\star$. In practice $\theta_\star$ is unknown and one only has access
to the observation sub-sequence $(Y_t)_{t\geq0}$, where, by convention, the
initial observation time is taken to be 0. In this context, the initial pdf
$\nu$ of $X_0$ is arbitrary and $\ell_{\nu,\theta}$, $p_{\nu,\theta}$,
$\P_{\nu,\theta}$ and $\E_{\nu,\theta}$ are used to denote, respectively, the
likelihood, the joint density of the states and observations, the probability,
and, the expectation under the non-stationary model that has initial pdf $\nu$
and $\theta$ as parameter. Hence, $\P_{\theta}$ refers to the probability under
$\theta$ of the stationary HMM process $(X_t,Y_t)_{t\in\mathbb{Z}}$ while
$\P_{\nu,\theta}$ denotes the probability of the non-stationary process
$(X_t,Y_t)_{t\in\mathbb{N}}$ which is started with initial distribution $\nu$
for $X_0$. Note that $\nu$ itself is not considered as a model parameter as it
cannot be estimated consistently from a single trajectory (see Chapters 10 and
12 of \citealp{cappe:moulines:ryden:2005} for further discussion of this
issue). The state transition matrix and state conditional pdf (probability
density function of $Y_t$ given $X_t$) that characterize the HMM are denoted,
respectively, by $q_\theta(x,x')$ and $g_\theta(x,y)$.

To make the EM recursion explicit, a key requirement is that the model belongs to an exponential family. In the following, we will thus make the following assumptions.

\begin{assumption} \
\begin{description}
\item[(i) Exponential Family] 
\begin{equation}
  p_{\theta}(x_t, y_t| x_{t-1}) = h(x_t,y_t) \exp\left(\langle \psi(\theta), s(x_{t-1},x_t,y_t) \rangle - A(\theta)\right) \eqsp ,
  \label{eq:exon}
\end{equation}
where $\langle \cdot \rangle$ denotes the scalar product, $s : (x',x,y) \in
\mathcal{X}^2 \times \mathcal{Y} \mapsto s(x',x,y) \in \mathcal{S}$ is the
vector of complete-data sufficient statistics, $\psi(\cdot)$ is the
(non-necessarily invertible) function that maps $\theta$ to the natural
parameterization and $A(\cdot)$ is the log-partition function.
\item[(ii) Explicit M-Step] For all $S \in \mathcal{S}$, the
  complete-data maximum likelihood equation
\[
  \nabla_\theta \psi(\theta) S - \nabla_\theta A(\theta) = 0 \eqsp ,
\]
where $\nabla_\theta$ denotes the gradient, has a unique solution denoted by $\bar{\theta}(S)$.
\end{description}
\label{hyp:main} 
\end{assumption}

Assumption~\ref{hyp:main}--(ii) states that the function $\bar{\theta} : S \in \mathcal{S} \mapsto
\bar{\theta}(S) \in \Theta$ that returns the complete-data maximum likelihood estimator
corresponding to any feasible value of the sufficient statistics is available in
closed-form. Note that the form used in~\eqref{eq:exon} is in fact slightly more general than the HMM
case. Indeed, if both $q_{\theta}$ and $g_{\theta}$ belong to exponential families:
\begin{align}
  & q_\theta(x',x) = h^q(x',x) \exp\left(\langle \psi^q(\theta), s^q(x',x) \rangle - A^q(\theta)\right) \eqsp , \nonumber \\
  & g_\theta(x,y) = h^g(x,y) \exp\left(\langle \psi^g(\theta), s^g(x,y) \rangle - A^g(\theta)\right) \eqsp ,
  \label{eq:expon:detail}
\end{align}
we then have $A(\theta) = A^q(\theta) + A^g(\theta)$,
\begin{equation*}
  \psi(\theta) =
  \begin{pmatrix}
    \psi^q(\theta) \\
    \psi^g(\theta)
  \end{pmatrix} \,  \text{and} \, s(x',x,y) =
  \begin{pmatrix}
    s^q(x',x) \\
    s^g(x,y)
  \end{pmatrix} \eqsp .
\end{equation*}
We will not make use of this specific structure when describing the algorithm
and we thus stick to the concise representation of~\eqref{eq:exon}. Sections
and \ref{sec:complex} and \ref{sec:gausHMM} provide more details on the
nature of the function $s$ in specific examples of HMMs.

Under Assumption~\ref{hyp:main}, the $k$-th iteration of the usual EM algorithm applied to observations $Y_{0:n}=(Y_0,\dots,Y_n)$ takes the following familiar form:
\begin{description}
\item[E-Step] Compute
\begin{equation} 
 S_{k+1} = \frac{1}{n}
\E_{\nu,\theta_k}\left[\left. \sum_{t=1}^n s(X_{t-1},X_t,Y_t) \right| Y_{0:n} \right] \eqsp .
\label{eq:EM:E-Step}
\end{equation}
\item[M-Step] Update the parameter estimate to $\theta_{k+1} = \bar{\theta}(S_{k+1})$.
\end{description}
To avoid unnecessary notational complexity, we have omitted in~\eqref{eq:EM:E-Step} the initial term $\frac1n \log p_{\nu,\theta}(x_0, y_0)$ from the normalized complete-data log-likelihood. The influence of this term is vanishing with $n$ and it is not necessary to take it into account for online estimation.

\subsection{Recursive Form of Smoothing}
We briefly recall here a key ingredient of the proposed algorithm which makes it possible to compute recursively the normalized sum $\frac{1}{n}
\E_{\nu,\theta}\left[\sum_{t=1}^n \left. s(X_{t-1},X_t,Y_t) \right| Y_{0:n} \right]$ by use of an auxiliary recursion. Curiously, this idea which dates back to, at least, \cite{zeitouni:dembo:1988} and has been extensively studied by \cite{elliott:aggoun:moore:1995} remains largely under-exploited (see discussion in Chapter 4 of \citealp{cappe:moulines:ryden:2005}).

In addition to the usual filter,
\begin{equation}
  \phi_{n,\nu,\theta}(x) = \P_{\nu,\theta}\left(\left.X_n = x \right| Y_{0:n}\right) \eqsp ,
\end{equation}
define the following intermediate quantity
\begin{equation}
  \rho_{n,\nu,\theta}(x) = \frac1n \E_{\nu,\theta}\left[\left. \sum_{t=1}^n s(X_{t-1},X_t,Y_t) \right| Y_{0:n}, X_n = x \right] \eqsp. \label{eq:stat:def}
\end{equation}
Obviously, these two quantities allow the computation of the sum of interest as\[
 \sum_{x\in\mathcal{X}} \phi_{n,\nu,\theta}(x) \rho_{n,\nu,\theta}(x) =
\frac1n \E_{\nu,\theta}\left[\left. \sum_{t=1}^n s(X_{t-1},X_t,Y_t) \right| Y_{0:n} \right] \eqsp .
\]
The appeal of
this decomposition is that $\phi_{n,\nu,\theta}$ and $\rho_{n,\nu,\theta}$ can be updated recursively according to the following proposition.

\begin{proposition}
  \label{prop:recursive}
\begin{description}
\item[Initialization] For $x \in \mathcal{X}$, set
  \begin{align*}
    & \phi_{0,\nu,\theta}(x) = \frac{\nu(x) g_\theta(x,Y_0)}{\sum_{x'\in\mathcal{X}} \nu(x') g_\theta(x',Y_0)} \eqsp ,\\
    & \rho_{0,\nu,\theta}(x) = 0 \eqsp .
  \end{align*}
\item[Recursion] For $n \geq 0$ and $x \in \mathcal{X}$, it holds that
  \begin{align}
   & \phi_{n+1,\nu,\theta}(x) = \frac{\sum_{x'\in\mathcal{X}} \phi_{n,\nu,\theta}(x')q_\theta(x',x)g_\theta(x,Y_{n+1})}{\sum_{x',x''\in\mathcal{X}^2} \phi_{n,\nu,\theta}(x')q_\theta(x',x'')g_\theta(x'',Y_{n+1})} \label{eq:filter:update} \\
   & \rho_{n+1,\nu,\theta}(x) = \sum_{x'\in\mathcal{X}} \bigg\{\frac{1}{n+1} s(x',x,Y_{n+1}) \nonumber \\
   & \qquad \qquad \qquad \qquad \qquad + \left(1-\frac{1}{n+1}\right)\rho_{n,\nu,\theta}(x') \bigg\} \frac{\phi_{n,\nu,\theta}(x') q_\theta(x', x)}{\sum_{x''\in\mathcal{X}} \phi_{n,\nu,\theta}(x'') q_\theta(x'', x)} \eqsp .
   \label{eq:stat:update}
\end{align}
\end{description}
\end{proposition}

In Proposition~\ref{prop:recursive} above, the rightmost term in~\eqref{eq:stat:update},
\[
  r_{n+1,\nu,\theta}(x'|x) =  \frac{\phi_{n,\nu,\theta}(x') q_\theta(x', x)}{\sum_{x''\in\mathcal{X}} \phi_{n,\nu,\theta}(x'') q_\theta(x'', x)} \eqsp ,
\]
corresponds to the \emph{backward retrospective probability}
$\P_{\nu,\theta}(X_n=x'|X_{n+1}=x,Y_{0:n})$, which does not depend on the newly
available observation $Y_{n+1}$. The main argument in proving
Proposition~\ref{prop:recursive} is to check that
\begin{multline*}
   \P_{\nu,\theta}(X_t=x_t, X_{t+1} = x_{t+1}|X_{n+1}=x_{n+1},Y_{0:n+1}) = \\
   \sum_{x_n\in\mathcal{X}}  \P_{\nu,\theta}(X_t=x_t, X_{t+1} = x_{t+1}|X_{n}=x_{n},Y_{0:n}) 
  \P_{\nu,\theta}(X_n=x_n|X_{n+1}=x_{n+1},Y_{0:n}) \eqsp ,
\end{multline*}
for all indices $0 \leq t \leq n-1$ which implies the claimed result by summation on $t$.

\subsection{Online EM Algorithm}
\label{sec:online}
Proposition~\ref{prop:recursive} constitutes a recursive rewriting of the computation required to carry out the E-step in the batch EM algorithm. By analogy with the case of independent observations studied in \cite{cappe:moulines:2009},
the proposed online EM algorithm for HMMs takes the following form.

\begin{algorithm}
\label{alg:generic}
Chose a decreasing sequence $(\gamma_n)_{n\geq 1}$ of step-sizes, which satisfy the usual stochastic approximation requirement that $\sum_{n\geq 1} \gamma_n = \infty$ and $\sum_{n\geq 1} \gamma_n^2 < \infty$. Also select a parameter initialization $\hat{\theta}_0$ and a minimal number of observations $n_{\min}$ required before performing the first parameter update.
\begin{description}
\item[Initialization] Compute, for $x \in \mathcal{X}$,
  \begin{align*}
    & \hat{\phi}_0(x) = \frac{\nu(x) g_{\hat{\theta}_0}\!(x,Y_0)}{\sum_{x'\in\mathcal{X}} \nu(x') g_{\hat{\theta}_0}\!(x',Y_0)} \eqsp ,\\
    & \hat{\rho}_0(x) = 0 \eqsp .
  \end{align*}
\item[Recursion] For $n \geq 0$,

  Compute, for $x \in \mathcal{X}$,
  \begin{align}
   & \hat{\phi}_{n+1}(x) = \frac{\sum_{x'\in\mathcal{X}} \hat{\phi}_{n}(x')q_{\hat{\theta}_n}\!(x',x)g_{\hat{\theta}_n}\!(x,Y_{n+1})}{\sum_{x',x''\in\mathcal{X}^2} \hat{\phi}_{n}(x')q_{\hat{\theta}_n}\!(x',x'')g_{\hat{\theta}_n}\!(x'',Y_{n+1})} \eqsp, \label{eq:stat:filter:online} \\
   & \hat{\rho}_{n+1}(x) = \sum_{x'\in\mathcal{X}} \left\{ \gamma_{n+1} s(x',x,Y_{n+1}) + (1-\gamma_{n+1}) \hat{\rho}_{n}(x') \right\} \frac{\hat{\phi}_{n}(x') q_{\hat{\theta}_n}\!(x', x)}{\sum_{x''\in\mathcal{X}} \hat{\phi}_{n}(x'') q_{\hat{\theta}_n}\!(x'', x)} \eqsp.
   \label{eq:stat:update:online}
\end{align}

  If $n \geq n_{\min}$, update the parameter according to
  \begin{equation}
    \hat{\theta}_{n+1} = \bar{\theta}\left( \sum_{x\in\mathcal{X}} \hat{\rho}_{n+1}(x) \hat{\phi}_{n+1}(x) \right) \eqsp, 
  \label{eq:onlineEM-M}
  \end{equation}
  otherwise, set $\hat{\theta}_{n+1} = \hat{\theta}_{n}$.
\end{description}
\end{algorithm}

In Algorithm~\ref{alg:generic}, the role of $n_{\min}$ is only to guarantee
that the M-step update is numerically well-behaved as it is well-known that in
most models the maximum likelihood estimation equation is degenerate for very
small numbers of observations and, hence, the function $\bar{\theta}$ may not be
properly defined. For this purpose, a very small value of $n_{\min}$ is usually
sufficient (for instance, $n_{\min}=20$ is used in the simulations of Section
\ref{sec:exper}).

\section{Discussion}
\label{sec:discuss}

We first discuss connections of the proposed algorithm with earlier works before discussing its
numerical complexity in more details.

Quite obviously, Algorithm~\ref{alg:generic} is intended to generalize the
online EM algorithm of \cite{cappe:moulines:2009} to the case of dependent
observations. In the case of independent observations,
\eqref{eq:stat:filter:online} and \eqref{eq:stat:update:online} reduces to
a simpler recursion of the form:
\begin{equation}
  \hat{S}_{n+1} = \gamma_{n+1} \E_{\bar{\theta}(\hat{S}_n)}\left[\left. s(X_{n+1},Y_{n+1})\right|Y_{n+1}\right] + (1-\gamma_{n+1})\hat{S}_{n}  \eqsp ,
  \label{eq:onlineEM-iid}
\end{equation}
which can be analyzed using the arguments developed for the analysis of stochastic approximation
(or Robbins-Monro) schemes under Markovian random perturbations. The HMM framework however implies
several key differences. First, in HMMs it is necessary to maintain an approximate filter
$\hat{\phi}_{n}$ through~(\ref{eq:stat:filter:online}) which, hopefully, becomes an acceptable
proxy for $\phi_{n,\nu,\hat{\theta}_n}$ when approaching convergence. More importantly, as it is no
more possible to compute the conditional expectation of the complete-data sufficient statistics
online, we require the auxiliary recursion of~(\ref{eq:stat:update:online}). Although directly
inspired by the exact recursive smoothing formula of~(\ref{eq:stat:update}), this auxiliary update
cannot be put in the usual stochastic approximation form. And thus, in contrast to the online EM
algorithm of \cite{cappe:moulines:2009}, Algorithm~\ref{alg:generic} cannot be analyzed using
off-the-shelf mathematical arguments. Both the need to maintain an estimate of the filter and the
presence of the backward retrospective probability in~(\ref{eq:stat:update:online}) constitute
significant departures from the usual stochastic approximation framework. Thus a complete analysis
of the convergence of Algorithm~\ref{alg:generic} is a challenging mathematical problem and we
provide in Section~\ref{sec:results} below important ---though limited--- arguments in that direction.

Algorithm~\ref{alg:generic} is related to the EM-based approach suggested in Eq. (51) of
\cite{kantas_doucet_singh_maciejowski_09} which however lacks the idea of a recursive
implementation of smoothing (in this work, the authors propose to use a particle filtering method
to approximate the smoothing functional). Chapters 4 and 5 of \cite{ford_98-thesis} and
\cite{elliott_ford_moore_02} consider the use of Proposition~\ref{prop:recursive} for online
estimation of the parameter of a Markov chain observed in additive noise ---an example that we will
discuss in more detail in Section~\ref{sec:gausHMM}--- and for a linear state-space model with
known noise characteristics. \cite{ford_98-thesis} however does not acknowledge the generality of
the approach and the role of complete-data sufficient statistics and is constrained to the choice
of $\gamma_n = n^{-1}$, which in simulations gives poor performances compared to alternative choices of
step-sizes (see Section~\ref{sec:gausHMM} below). To the best of our knowledge, the first instance
of a particular case of Algorithm~\ref{alg:generic} appears in the work of
\cite{mongillo:deneve:2008} who considered the specific case of finite-valued
observations. \cite{mongillo:deneve:2008} however entirely rediscovered
Proposition~\ref{prop:recursive} under an equivalent form discussed below and thus failed to
identify some of the general principles underpinning Algorithm~\ref{alg:generic}.

\subsection{Comparison with the Algorithm of \cite{mongillo:deneve:2008}}
This section is devoted to a more detailed analysis of the difference between
Algorithm~\ref{alg:generic} applied to the case of finite-valued observations and the actual
proposal of \cite{mongillo:deneve:2008}. This difference is not significant from a
practical point of view but is important for the understanding of the behavior of the algorithm.

\cite{mongillo:deneve:2008} considered the particular case of finite valued
HMMs in which the observations $(Y_t)_{t\geq1}$ also take their values in a finite
set $\mathcal{Y}$. In such a situation, it is easily checked that for any
parameterization of the model, the complete-data
sufficient statistics may be chosen as $s(X_{t-1},X_t,Y_{t}) =
(\1\{X_{t-1}=i,X_t=j,Y_{t}=k\})_{(i,j,k) \in \mathcal{X}^2 \times
  \mathcal{Y}}$. The recursion derived by \cite{mongillo:deneve:2008} for this
case is based on recursively updating the product $\tau_{n,\nu,\theta}(x) =
\phi_{n,\nu,\theta}(x)\rho_{n,\nu,\theta}(x)$ rather than
$\rho_{n,\nu,\theta}(x)$. The probabilistic interpretation of the new term
$\tau_{n,\nu,\theta}(x)$ is $\E_{\nu,\theta}\left[\left. \left(\sum_{t=1}^n
      s(X_{t-1},X_t,Y_t)\right) \1\{X_n = x \} \right| Y_{0:n}\right]$. By
multiplying \eqref{eq:stat:update:online} by $\hat{\phi}_{n+1}(x)$ and
using~\eqref{eq:stat:filter:online}, one obtains the following online
update
  \begin{multline}
    \hat{\tau}_{n+1}(x) = \gamma_{n+1} \sum_{x'\in\mathcal{X}} s(x',x,Y_{n+1}) \frac{\hat{\phi}_{n}(x') q_{\hat{\theta}_n}\!(x', x) g_{\hat{\theta}_n}\!(x',Y_{n+1})}{\sum_{x',x''\in\mathcal{X}^2} \hat{\phi}_{n}(x') q_{\hat{\theta}_n}\!(x', x'') g_{\hat{\theta}_n}\!(x'',Y_{n+1})} \\
    + (1-\gamma_{n+1})  \sum_{x'\in\mathcal{X}} \hat{\tau}_{n}(x') \frac{q_{\hat{\theta}_n}\!(x', x) g_{\hat{\theta}_n}\!(x',Y_{n+1})}{\sum_{x',x''\in\mathcal{X}^2} \hat{\phi}_{n}(x') q_{\hat{\theta}_n}\!(x', x'') g_{\hat{\theta}_n}\!(x'',Y_{n+1})} \eqsp ,
    \label{eq:update:online:joint}
  \end{multline}
which coincides with Eqs. (15)-(16) of \cite{mongillo:deneve:2008}, for the particular choice of
complete-data sufficient statistics discussed above.

Of course, using either~(\ref{eq:stat:filter:online}) and~\eqref{eq:stat:update:online} or
\eqref{eq:update:online:joint} is practically equivalent. But the results of
Section~\ref{sec:results} indicate that $\hat{\phi}_n$ and $\hat{\rho}_n$ have very different
limiting behaviors: the auxiliary recursion~(\ref{eq:stat:update:online}) should converge to a
fixed deterministic limit while the same is not true for the approximate filtering
recursion~(\ref{eq:stat:filter:online}). Hence the use of decreasing step-sizes
for~(\ref{eq:stat:update:online}) only is justifiable while its use
in~\eqref{eq:update:online:joint} is less natural (and/or potentially misleading) as
$\hat{\tau}_{n}$ should not be expected to converge to a deterministic limit.

Regarding the choice of the step-size, \cite{mongillo:deneve:2008} consider the cases where,
either, the step-size $\gamma_n$ is small but non-decreasing, which may be useful for tracking
potential changes but is not sufficient to guarantee the consistency of the approach. The other
option mentioned by \cite{mongillo:deneve:2008} is to use $\gamma_n = n^{-1}$ by analogy with the
work of \cite{neal:hinton:1999} and the case of the batch EM algorithm. The range of
step-sizes mentioned in Algorithm~\ref{alg:generic} is chosen in reference to the theory of
stochastic approximation and by analogy with \cite{cappe:moulines:2009}. As will be illustrated
below, in the numerical simulations of Section~\ref{sec:gausHMM}, the choice of $\gamma_n = n^{-1}$
should definitely be avoided for HMMs and we instead recommend using step-sizes of the form $\gamma_n = n^{-\alpha}$ with $\alpha$ in the interval (0.5,0.8), possibly combined with
Polyak-Ruppert averaging (see Section~\ref{sec:exper} below).

\subsection{Implementation and Numerical Complexity}
\label{sec:complex}
Regarding the numerical complexity of Algorithm~\ref{alg:generic}, observe that in the case
considered by \cite{mongillo:deneve:2008} where $s(X_{t-1},X_t,Y_{t}) =
(\1\{X_{t-1}=i,X_t=j,Y_{t}=k\})_{(i,j,k) \in \mathcal{X}^2 \times \mathcal{Y}}$,
$s(X_{t-1},X_t,Y_{t})$ is a vector of dimension $|\mathcal{X}|^2 \times |\mathcal{Y}|$ (where
$|\cdot|$ denotes the cardinal of the set). Thus, the numerical complexity
of~\eqref{eq:update:online:joint} is of order $|\mathcal{X}|^4 \times |\mathcal{Y}|$ per
observation. For this case, it is indeed possible to bring down the numerical complexity to the
order of $|\mathcal{X}|^4$ + $|\mathcal{X}|^3 \times |\mathcal{Y}|$ operations by updating
separately the terms corresponding to the two statistics $(\1\{X_{t-1}=i,X_t=j\})_{(i,j) \in
  \mathcal{X}^2}$ and $(\1\{X_t=j,Y_{t}=k\})_{(j,k) \in \mathcal{X} \times \mathcal{Y}}$ (see the
example considered in Section~\ref{sec:gausHMM} for more details). Interestingly, the numerical complexity
of the batch EM algorithm for this model, when implemented using traditional forward-backward
smoothing \citep{rabiner:1989}, is of the order of $(|\mathcal{X}|^2$ + $|\mathcal{X}| \times
|\mathcal{Y}|)$ per observation and per iteration of the EM algorithm. The comparison is
not directly meaningful as the batch EM algorithm does necessitate several iterations to converge
(see numerical illustrations in Section~\ref{sec:exper}). The scaling of the
numerical complexity of the online-EM algorithm with respect to $|\mathcal{X}|$ can constitute an hindrance in
models with a large number of states. This being said, the complexity of online gradient-based
approaches, is equivalent as the main burden comes from the necessity of updating, via a recursion
related to~\eqref{eq:stat:update:online}, one coordinate of the gradient for each of the pairs
$(x,x')\in\mathcal{X}^2$ of state values (see, e.g., \citealp{legland:mevel:1997}). When the transition
matrix is structured ---i.e., parametered by a low dimensional parameter rather than by all its
individual entries---, the numerical cost of implementing the online EM approach is
reduced to an order of the number of parameters times $|\mathcal{X}|^2$.

\section{Some Results on Convergence}
\label{sec:results}
In this section, we provide two important elements regarding Algorithm~\ref{alg:generic}. The
first is a generalization of the argument of \cite{cappe:moulines:2009} that makes it possible to
identify a deterministic limit for the EM update as the number of observations tends to
infinity. Interestingly, the form of this limit is non trivial and substantially different from the
case of independent observations. The second result
pertains to the limiting behavior of the auxiliary quantity $\hat{\rho}_{n}(x)$
of~\eqref{eq:onlineEM-M} that is instrumental in Algorithm~\ref{alg:generic}. As discussed in the
previous section, a complete analysis of the convergence of Algorithm~\ref{alg:generic} is still
lacking but it is possible to show that when the parameter is frozen (i.e. when the M-step update
of (\ref{eq:onlineEM-M}) is inhibited), $\hat{\rho}_{n}$ converges to a deterministic quantity. This limit \emph{does not depend
  on} $x$ (or, in other words, $|\hat{\rho}_{n}(x)-\hat{\rho}_{n}(x')| \to 0$)
and is related to the limiting EM mapping. This result, although limited, is very
important to understand the nature of the asymptotic attractors of
Algorithm~\ref{alg:generic}. In this section, we will work under the assumptions of
\cite{douc:moulines:ryden:2004} which guarantee the asymptotic normality of the MLE, adapted to the
(simplest) case of a finite state space $\mathcal{X}$.

We start by a closer inspection of the limiting behavior of the normalized score function (gradient of the log-likelihood). Under suitable assumptions (see below), the normalized HMM log-likelihood $\frac{1}{n} \log \ell_{\nu,\theta}(Y_0,\dots,Y_n)$
converges, $\P_{\theta_\star}$ almost surely and in $L^1$, to the limiting contrast
\begin{equation}
  c_{\theta_\star}(\theta) = \E_{\theta_\star}[\log \ell_\theta(Y_0|Y_{-\infty:-1})] \eqsp .
  \label{eq:HMM:contrast}
\end{equation}
The same is true for the normalized score $\frac{1}{n} \nabla_{\theta} \log
\ell_{\nu,\theta}(Y_0,\dots,Y_n)$ which converges to $\nabla_\theta
c_{\theta_\star}(\theta)$. Such consistency results have been established, under various assumptions, by (among others) \cite{baum:petrie:1966,bickel:ritov:ryden:1998,douc:moulines:ryden:2004}. Now, thanks to Fisher identity, for all $n$,
\begin{multline}
  \frac{1}{n} \nabla_{\theta} \log \ell_{\nu,\theta}(Y_0,\dots,Y_n) = \frac{1}{n} \E_{\nu,\theta}\left[\left. \sum_{t=1}^n \nabla_\theta \log p_{\theta}(X_t, Y_t| X_{t-1}) \right| Y_{0:n} \right] \\
  + \frac{1}{n} \E_{\nu,\theta}\left[\left. \nabla_{\theta} \log p_{\nu,\theta}(X_0, Y_0) \right| Y_{0:n} \right] \eqsp .
  \label{eq:Fisher-HMM}
\end{multline}
As already discussed, the last term on the r.h.s., whose influence is vanishing with increasing values of $n$, can be ignored. Hence, the consistency result for the score function combined with \eqref{eq:Fisher-HMM} implies that $\frac{1}{n} \E_{\nu,\theta}\left[\left. \sum_{t=1}^n \nabla_\theta \log p_{\theta}(X_t, Y_t| X_{t-1}) \right| Y_{0:n} \right]$ converges $\P_{\theta_\star}$ almost surely to  $\nabla_\theta c_{\theta_\star}(\theta)$, the gradient of the limiting contrast. Now using the exponential family representation in~(\ref{eq:exon}), the non-vanishing term in the r.h.s. of~\eqref{eq:Fisher-HMM} may
rewritten as
\begin{multline}
  \frac{1}{n} \E_{\nu,\theta}\left[\left. \sum_{t=1}^n \nabla_\theta \log p_{\theta}(X_t, Y_t| X_{t-1}) \right| Y_{0:n} \right] = \\
   \nabla_\theta\psi(\theta) \left\{ \frac{1}{n} \E_{\nu,\theta}\left[\left. \sum_{t=1}^n s(X_{t-1},X_t,Y_t) \right| Y_{0:n} \right] \right\} - \nabla_\theta A(\theta) \eqsp .
\label{eq:gradient_expform}
\end{multline}
The following theorem defines the limiting behavior of the r.h.s. of the above equation, and thus,
the limiting behavior of the EM update for HMMs (see Appendix~\ref{appendix} for the corresponding proofs).

\begin{theorem}
  \label{thm:limitingEM}
  In addition to Assumption~\ref{hyp:main}, assume that \emph{(i)} $\mathcal{X}$ is a finite set; \emph{(ii)} the parameter space $\Theta$ is compact and $\theta_\star$ lies in the interior of $\Theta$; \emph{(iii)} the transition matrix is such
  that $q_\theta(x,x') \geq \epsilon > 0$ for all $\theta \in \Theta$; \emph{(iv)}
  $\sup_\theta\sup_y \bar{g}_\theta(y) < \infty$ and $\E_{\theta_\star}\left[\left|\log\inf_\theta
      \bar{g}_\theta(Y_0)\right|\right] < \infty$, where $ \bar{g}_\theta(y) = \sum_x
  g_\theta(x,y)$;  and, \emph{(v)} $\psi_q, A_q, \psi_g, A_g$ in~\eqref{eq:expon:detail}
  are continuously differentiable functions on the interior of $\Theta$. Then the following properties hold.

  \begin{enumerate}
  \item [(i)]   \begin{equation}
    \frac{1}{n} \E_{\nu,\theta}\left[\left. \sum_{t=1}^n s(X_{t-1},X_t,Y_t) \right| Y_{0:n} \right] \longrightarrow \E_{\theta_\star}\left(\E_\theta\left[\left. s(X_{-1},X_0,Y_0) \right| Y_{-\infty:\infty} \right] \right) \eqsp, \eqsp \text{$\P_{\theta_\star}$ a.s.}
  \label{eq:limitEM}
  \end{equation}
  \item [(ii)] The fixed points of the \emph{limiting EM algorithm}
  \begin{equation}
    \theta_{k+1} = \bar{\theta}\left\{ \E_{\theta_\star}\left( \E_{\theta_k}\left[\left. s(X_{-1},X_0,Y_0) \right| Y_{-\infty:\infty} \right] \right) \right\}
    \label{eq:limingEM-Hmm}    
  \end{equation}
  are the stationary points of the limiting likelihood contrast $c_{\theta_\star}(\theta)$.
  \end{enumerate}
\end{theorem}

Theorem~\ref{thm:limitingEM} provides an interpretation of the limiting form of the classical
EM algorithm when used on very long sequences of observations. In the case of HMMs, this form is
quite complicated and it is interesting to compare it to the case of independent observations
investigated by \cite{cappe:moulines:2009}. If the observations are assumed to be independent, the
law of large number implies the convergence of the normalized log-likelihood to
$c_{\theta_\star}(\theta) = \E_{\theta_\star}[\log \ell_\theta(Y_0)]$. In this case, it is obvious that
maximizing $c_{\theta_\star}(\theta)$ is equivalent to minimizing the Kullback-Leibler divergence
$D(\ell_{\theta_\star}|\ell_\theta) = \int
\log\frac{\ell_{\theta_\star}(y)}{\ell_\theta(y)}\ell_{\theta_\star}(y) dy$. In the HMM case, $D(\ell_{\theta_\star}|\ell_\theta)$ needs to be replaced by the
expression of $c_{\theta_\star}(\theta)$ given in~(\ref{eq:HMM:contrast}), which is a
consequence of the tower property of conditional expectation and of the forgetting property of the
filter \citep{douc:moulines:ryden:2004}. However, in contrast to the case of independent
observations, it is no more straightforward to provide an explicit expression for the gradient of
$c_{\theta_\star}(\theta)$. The key idea here is the use of Fisher identity
in~(\ref{eq:Fisher-HMM}) which yields the limiting term found in the
r.h.s. of~\eqref{eq:limitEM}. The use of Fisher identity also explains the particular form of
conditioning found in~\eqref{eq:limitEM}, which involves both the infinite future and past of the
trajectory. This expression also suggests that being able to compute or approximate smoothed
functionals of the state recursively is indeed a key requirement for online
estimation in HMMs.

The next result shows that under parameter freeze, the online EM update
equation~(\ref{eq:stat:filter:online}) converges to a deterministic constant function, equal to the
r.h.s. of~\eqref{eq:limitEM}.

\begin{corollary}
  \label{corr:rho_freeze}
  Under the assumptions of Theorem~\ref{thm:limitingEM}, Algorithm~\ref{alg:generic} used
  without~\eqref{eq:onlineEM-M} ---that is, with $\hat{\theta}_n$ equal to a fixed value $\theta$---
  satisfies
  \[
    \hat{\rho}_n(x) \longrightarrow \E_{\theta_\star}\left(\E_\theta\left[\left. s(X_{-1},X_0,Y_0) \right| Y_{-\infty:\infty} \right] \right) \eqsp, \eqsp \text{$\P_{\theta_\star}$ a.s., for all $x \in \mathcal{X}$.}
  \]
\end{corollary}

Corollary~\ref{corr:rho_freeze} shows that under parameter freeze the auxiliary quantity
$\hat{\rho}_n(x)$ converges to a constant limit that does not depend on $x$ and that is equal to
the limit obtained in~\eqref{eq:limitEM}. Note that since $\hat{\phi}_n(x)$ is a probability on
$\mathcal{X}$, this implies that $\bar{\theta}( \sum_{x\in\mathcal{X}} \hat{\rho}_{n+1}(x)
  \hat{\phi}_{n+1}(x) )$ tends to $\bar{\theta}\left\{ \E_{\theta_\star}\left(
    \E_{\theta}\left[\left. s(X_{-1},X_0,Y_0) \right| Y_{-\infty:\infty} \right] \right)
\right\}$. This result together with Theorem~\ref{thm:limitingEM}--(ii) suggests that Algorithm~\ref{alg:generic} can only be stable at the
stationary points of the limiting contrast $c_{\theta_\star}(\theta)$. The argument is only
heuristic at this stage as Corollary~\ref{corr:rho_freeze} is obtained under the artificial
assumption that the evolution of the parameter is frozen. Corollary~\ref{corr:rho_freeze}
however highlights an essential characteristic of Algorithm~\ref{alg:generic}: in contrast to
$\hat{\phi}_n(x)$ which does approximate the infinite past filter and hence varies with each
observation, $\hat{\rho}_n(x)$ converges to a deterministic limit that is independent of $x$. This
property, which can be verified in simulations, justifies the form of
Algorithm~\ref{alg:generic} and in particular the use of a stochastic approximation type of update
for $\hat{\rho}_n(x)$ only.

\section{Application to Gaussian HMMs}
\label{sec:gausHMM}
\subsection{HMM with Product Parameterization}
For the sake of concreteness, we consider in following the case where the state variables
$(X_t)$ take their values in the set $\{1,\dots,m\}$. In addition, assume that, as is
often the case in practise, the parameter $\theta$ may be split into two sub-components that
correspond, respectively, to the state transition matrix $q_\theta$ and to the state-conditional
densities $\{g_\theta(i,\cdot)\}_{1\leq i\leq m}$. In the fully discrete case considered in
\cite{mongillo:deneve:2008} for instance, the parameter $\theta$ consist of the transition matrices
$q_\theta$ and $g_\theta$ parametered by their respective entries, with the constraint that each
line of a transition matrix must sum to one. In the case of Gaussian HMMs used in speech
processing as well as many in other applications, the parameters are the state transition matrix
$q_\theta$ and the mean vector and covariance matrix associated with each of the $m$
state-conditional densities $\{g_\theta(i,\cdot)\}_{1\leq i\leq m}$ \citep{rabiner:1989}.

In such a model, there are two distinct types of EM complete-data sufficient statistics which give
rise to two separate forms of the auxiliary function $\rho_{n,\nu,\theta}$:
\begin{align}
  & \rho^{q}_{n,\nu,\theta}(i,j,k;\theta) = \frac1n \E_{\nu,\theta} \left[\left.\sum_{t=1}^n \1\{X_{t-1}=i,X_t=j\}\right|Y_{0:n}, X_n = k \right] \eqsp , \label{eq:stat_trans} \\
  & \rho^{g}_{n,\nu,\theta}(i,k;\theta) = \frac1n \E_{\nu,\theta} \left[\left.\sum_{t=0}^n \1\{X_t=i\}s(Y_t)\right|Y_{0:n}, X_n = k \right] \eqsp , \label{eq:stat_obs}
\end{align}
where the form of $s$ itself depend on the nature of the state-conditional distribution
$g_\theta(x,\cdot)$ ---see Gaussian example below. There's a slight difference between \eqref{eq:stat_obs} and \eqref{eq:stat:def}, which is that \eqref{eq:stat_obs} also incorporates the initial ($t=0$) conditional likelihood term, i.e., the contribution corresponding to the rightmost term on the r.h.s. of (\ref{eq:Fisher-HMM}). As noted earlier, this difference is minor and does not modify the long-term behavior of the algorithm.


With these notations, Eq.~\eqref{eq:stat:update:online} in Algorithm~\ref{alg:generic} is implemented as
\begin{align}
    & \hat{\rho}_{n+1}^{q}(i,j,k) = \gamma_{n+1} \delta(j-k) \hat{r}_{n+1}(i|j) + \left(1-\gamma_{n+1}\right) \sum_{k'=1}^m \hat{\rho}_{n}^{q}(i,j,k') \hat{r}_{n+1}(k'|k) \eqsp , \label{eq:SA_E-step:tran} \\
    & \hat{\rho}_{n+1}^{g}(i,k) =  \gamma_{n+1} \delta(i-k) s(Y_{n+1}) + \left(1-\gamma_{n+1}\right) \sum_{k'=1}^m \hat{\rho}_{n}^{g}(i,k') \hat{r}_{n+1}(k'|k) \eqsp , \label{eq:SA_E-step:obs}
\end{align}
where $\delta$ denotes the Kronecker delta (i.e., $\delta(i) = 0$ iff $i=0$) and the notation $\hat{r}_{n+1}(i|j)$ refers to the approximate retrospective conditional probability~:
\begin{equation}
  \hat{r}_{n+1}(i|j) = \frac{\hat{\phi}_n(i) q_{\hat{\theta}_n}\!(i, j)}{\sum_{i'=1}^m \hat{\phi}_n(i') q_{\hat{\theta}_n}\!(i', j)} \eqsp .
  \label{eq:retro}
\end{equation}

A complete iteration of the online algorithm involves the approximate filter update
\eqref{eq:stat:filter:online} and the statistics updates
\eqref{eq:SA_E-step:tran} and~\eqref{eq:SA_E-step:obs} followed by an application of the M-step
function $\bar{\theta}$ to $\hat{S}^{q}_{n+1}(i,j) = \sum_{k=1}^m \hat{\rho}^{q}_{n+1}(i,j,k)
\hat{\phi}_{n+1}(k)$ and $\hat{S}^{g}_{n+1}(i) = \sum_{k=1}^m
\hat{\rho}^{g}_{n+1}(i,k)\hat{\phi}_{n+1}(k)$. The form of the M-step depends on the exact nature
of $q_\theta$ and $g_\theta$. If the transition matrix $q_\theta$ is parametered simply by its entries,
the update is generic and is given by
\begin{equation}
  q_{\hat{\theta}_n}\!(i,j) = \bar{\theta}\left(\hat{S}^{q}_n\right) = \frac{\hat{S}^{q}_n(i,j)}{\sum_{j = 1}^m \hat{S}^{q}_n(i,j)} \eqsp .
  \label{eq:M-step:trans}
\end{equation}
For the update of the state-dependent parameters, one needs to be more specific and the form of the
equations depend on the choice of the state conditional density $g_\theta(x,\cdot)$. In the
multivariate Gaussian case, the function $s$ has to be chosen such that $s(y)$ consists of the three components
$\{1, y, y y^\ut\}$. The corresponding components of the approximated EM extended statistics are
denoted, respectively, by $\hat{S}_{n,0}^{g}, \hat{S}_{n,1}^{g}, \hat{S}_{n,2}^{g}$. If the state conditional Gaussian densities are parametered by their mean vectors,   $\mu_\theta(i)$, and
covariances matrices, $\Sigma_\theta(i)$, the M-step update is defined as
\begin{align}
  \mu_{\hat{\theta}_n}\!(i) & = \bar{\theta}\left(\hat{S}^{g}_{n,0},\hat{S}^{g}_{n,1}\right) =  \frac{\hat{S}^{g}_{n,1}(i)}{\hat{S}^{g}_{n,0}(i)} \label{eq:M-step:mean} \eqsp , \\
  \Sigma_{\hat{\theta}_n}\!(i) & = \bar{\theta}\left(\hat{S}^{g}_{n,0},\hat{S}^{g}_{n,1},\hat{S}^{g}_{n,2}\right) =
  \frac{\hat{S}^{g}_{n,2}(i)}{\hat{S}^{g}_{n,0}(i)} -  \mu_{\hat{\theta}_n}\!(i)\mu_{\hat{\theta}_n}^\ut\!(i)\label{eq:M-step:var} \eqsp .
\end{align}
The derivation of \eqref{eq:M-step:trans} and \eqref{eq:M-step:mean}--\eqref{eq:M-step:var} is straightforward but some more details are provided in the next section for a particular case of Gaussian HMM.

\subsection{Markov Chain Observed in Gaussian Noise}
In the numerical experiments described below, we consider the simple scalar model
\[
  Y_t = X_t + V_t \eqsp ,
\]
where $(V_t)$ is a scalar additive Gaussian noise of variance $\upsilon$ and $(X_t)$ is a Markov chain with transition matrix $q$, which takes its values in the set $\{\mu(1), \dots, \mu(m)\}$. Although simple, this model is already statistically challenging and is of some importance in several applications, in particular, as a basic model for ion channels data \citep{chung:moore:xia:premkumar:gage:1990} ---see also, e.g.,  \citep{roberts:ephraim:2008} for discussion of the continuous time version of the model as well \cite{degunst:kuensch:schouten:2001}, for an up-to-date account of models for ion channels. The parameter $\theta$ is comprised of the transition matrix $q$, the vector of means $\mu$ and the variance $\upsilon$.

In this case, the intermediate quantity of the batch EM algorithm may be written ---with constant terms omitted, as
\begin{equation}
   \sum_{i=1}^m \sum_{j=1}^m S^{q}_{n}(i,j) \log q(i,j) - \frac{1}{2\upsilon} \sum_{i=1}^m \left(S^{g}_{n,2}(i;\theta) - 2\mu(i) S^{g}_{n,1}(i) + \mu^2(i)  S^{g}_{n,0}(i) \right) \eqsp ,
  \label{eq:Q_EM:MCinNoise}
\end{equation}
where
\begin{align*}
 & S_n^q(i,j) = \frac1n \E_{\nu,\theta} \left[\left.\sum_{t=1}^n \1\{X_{t-1}=i, X_t=j\}\right|Y_{0:n}\right] \eqsp , \\
 & S^{g}_{n,0}(i) = \frac1n \E_{\nu,\theta} \left[\left.\sum_{t=0}^n \1\{X_t=i\}\right|Y_{0:n}\right] \eqsp , \\
 & S^{g}_{n,1}(i) = \frac1n \E_{\nu,\theta} \left[\left.\sum_{t=0}^n \1\{X_t=i\} Y_t\right|Y_{0:n}\right] \eqsp , \\
 & S^{g}_{n,2}(i) = \frac1n \E_{\nu,\theta} \left[\left.\sum_{t=0}^n \1\{X_t=i\} Y_t^2\right|Y_{0:n}\right] \eqsp .
\end{align*}
Maximization of~\eqref{eq:Q_EM:MCinNoise} with respect to $q_\theta$, $\mu_\theta$ and $v_\theta$ directly yields \eqref{eq:M-step:trans} as well as
\begin{align}
  \mu(i) & = \bar{\theta}\left(S^{g}_{n,0},S^{g}_{n,1}\right) =  \frac{S^{g}_{n,1}(i)}{S^{g}_{n,0}(i)} \eqsp , \label{eq:M-step:MCinNoise:mean} \\
  \upsilon & = \bar{\theta}\left(S^{g}_{n,0},S^{g}_{n,1},S^{g}_{n,2}\right) = \frac{\sum_{i=1}^m \left(S^{g}_{n,2}(i) -  \mu^2(i) S^{g}_{n,0}(i)\right)}{\sum_{i=1}^m S^{g}_{n,0}(i)} \label{eq:M-step:MCinNoise:var} \eqsp .
\end{align}
It is easily checked that, as usual, the M-step equations~\eqref{eq:M-step:trans}
and~\eqref{eq:M-step:MCinNoise:var} satisfy the constraints that $q$ be a stochastic
matrix and $\upsilon$ be non-negative.
Note that for this particular model, the use of the statistic $S^{g}_{n,2}$ could be avoided as it is only needed in the M-step under the form $\sum_{i=1}^m S^{g}_{n,2}(i)$, which is equal to $\frac1n \sum_{t=0}^n Y_t^2$.
Algorithm~\ref{alg:MC+noise} below recaps the complete online EM algorithm pertaining to this
example.

\begin{algorithm}[Online EM algorithm for noisily observed $m$-state Markov chain] \
\label{alg:MC+noise}
\begin{description}
\item[Initialization] \
  Select $\hat{\theta}_0$ and compute, for all $1 \leq i,j,k, \leq m$ and $0 \leq d \leq 2$,
  \begin{align*}
    & \hat{\phi}_{0}(k) = \frac{\nu(k)g_{\hat{\theta}_0}\!(k,Y_0)}{\sum_{k'=1}^m g_{\hat{\theta}_0}\!(k,Y_0)} \eqsp , \\
    & \hat{\rho}^{q}_{0}(i,j,k) = 0 \eqsp , \\
    & \hat{\rho}^{g}_{0,d}(i,k) = \delta(i-k) Y_0^d \eqsp .
  \end{align*}
\item[Recursion] For $n \geq 0$, and $1 \leq i,j,k, \leq m$, $0 \leq d \leq 2$, \
\begin{description}
\item[Approx. Filter Update] \
  \begin{equation*}
    \hat{\phi}_{n+1}(k) = \frac{\sum_{k'=1}^m \hat{\phi}_{n}(k')\hat{q}_{n}(k',k)g_{\hat{\theta}_n}\!(k,Y_{n+1})}{\sum_{k',k''=1}^m \hat{\phi}_{n}(k')\hat{q}_{n}(k',k'')g_{\hat{\theta}_n}\!(k'',Y_{n+1})} \eqsp ,
  \end{equation*}
  where $g_{\hat{\theta}_n}\!(k,y) =
\exp\left[-(y-\hat{\mu}_n(k))^2/2\hat{\upsilon}_n\right]$.
\item[Stochastic Approximation E-step] \
\begin{align*}
    & \hat{\rho}_{n+1}^{q}(i,j,k) = \gamma_{n+1} \delta(j-k) \hat{r}_{n+1}(i|j)
    + \left(1-\gamma_{n+1}\right) \sum_{k'=1}^m \hat{\rho}_{n}^{q}(i,j,k') \hat{r}_{n+1}(k'|k) \eqsp , \\
    & \hat{\rho}_{n+1,d}^{g}(i,k) =  \gamma_{n+1} \delta(i-k) Y_{n+1}^d
    + \left(1-\gamma_{n+1}\right) \sum_{k'=1}^m \hat{\rho}_{n,d}^{g}(i,k') \hat{r}_{n+1}(k'|k) \eqsp ,
\end{align*}
where $\hat{r}_{n+1}(i|j) = \hat{\phi}_n(i) \hat{q}_{n}(i, j) \big / \sum_{i'=1}^m \hat{\phi}_n(i') \hat{q}_{n}(i', j)$.
\item[M-step] If $n \geq n_{\min}$,
\begin{align*}
  & \hat{S}^{q}_{n+1}(i,j) = \sum_{k'=1}^m \hat{\rho}_{n+1}^{q}(i,j,k') \hat{\phi}_{n+1}(k') \eqsp , \\ 
  & \hat{q}_{n+1}(i,j) = \frac{\hat{S}^{q}_{n+1}(i,j)}{\sum_{j'=1}^m \hat{S}^{q}_{n+1}(i,j')} \eqsp , \\
  & \hat{S}^{g}_{n+1,d}(i) = \sum_{k'=1}^m \hat{\rho}_{n+1,d}^{g}(i,k') \hat{\phi}_{n+1}(k') \eqsp, \\
  & \hat{\mu}_{n+1}(i) = \frac{S^{g}_{n+1,1}(i)}{S^{g}_{n+1,0}(i)} \eqsp , \\
  & \hat{\upsilon}_{n+1} = \frac{\sum_{i'=1}^m \left(S^{g}_{n+1,2}(i') -  \hat{\mu}_{n+1}^2(i') S^{g}_{n+1,0}(i')\right)}{\sum_{i'=1}^m S^{g}_{n+1,0}(i')} \eqsp .
\end{align*}
\end{description}
\end{description}  
\end{algorithm}


\subsection{Numerical Experiments}
\label{sec:exper}
Algorithm~\ref{alg:MC+noise} is considered in the case of a two-state ($m=2$) model estimated from trajectories simulated from the model with parameters
\begin{align*}
  & q_{\star}(1,1) = 0.95 \eqsp , \eqsp \mu_\star(1) = 0 \eqsp , \\
  & q_{\star}(2,2) = 0.7 \eqsp , \eqsp \mu_\star(2) = 1 \eqsp , \\
  & \upsilon_\star = 0.5 \eqsp .
\end{align*}
With these parameters, state identification is a difficult task as the separation of the means corresponding to the two states is only 1.4 times the noise standard deviation. The optimal filter associated with the actual parameter does for instance misclassify the state (using Bayes' rule) in about 10.3\% of the cases. As will be seen below, this is reflected in slow convergence of the EM algorithm.

All estimation algorithms are systematically started from the initial values
\begin{align*}
  & q_{\star}(1,1) = 0.7 \eqsp , \eqsp \mu_\star(1) = -0.5 \eqsp , \\
  & q_{\star}(2,2) = 0.5 \eqsp , \eqsp \mu_\star(2) = 0.5 \eqsp , \\
  & \upsilon_\star = 2 \eqsp ,
\end{align*}
and run on 100 independent trajectories simulated from the model.

\begin{figure}[hbt]
  \centering
  \includegraphics[width=0.6\textwidth]{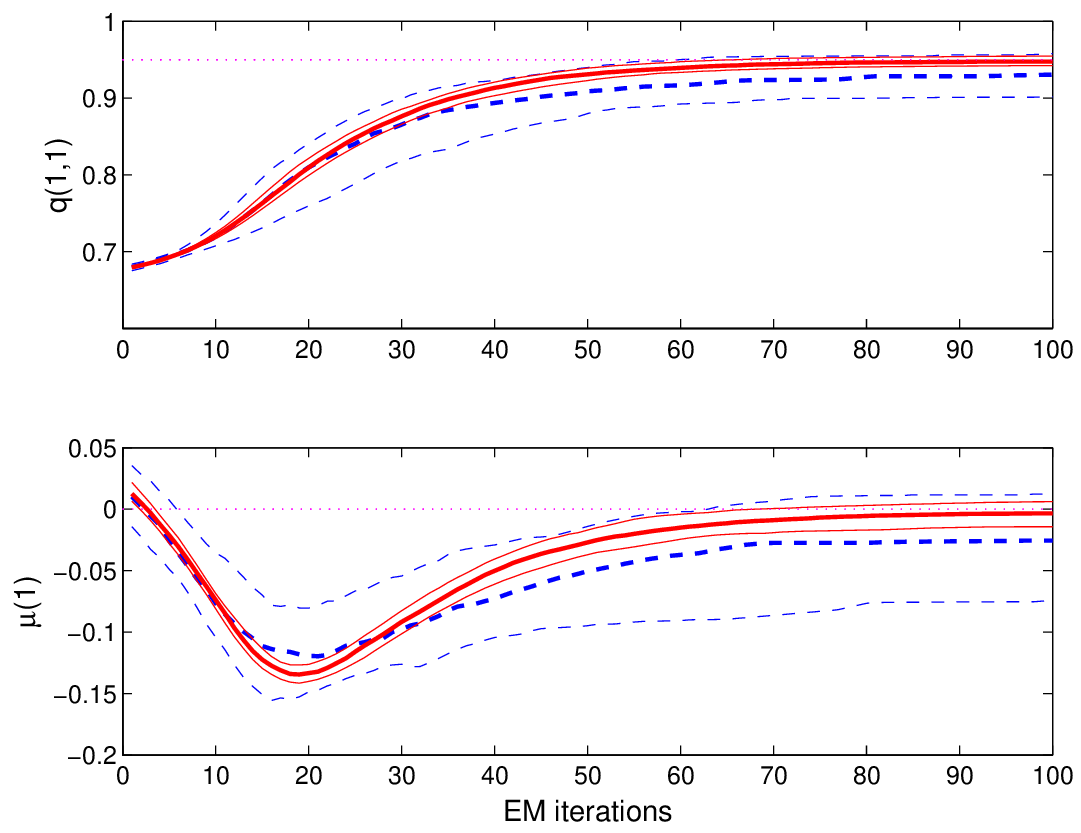}  
  \caption{Estimated values of $q(1,1)$ (top) and $\mu(1)$ (bottom) as a function of the number of batch EM iterations for $n=500$ (dotted lines) and $n=8000$ (solid lines) observations. The plot is based on 100 independent runs summarized by the median (bold central line) and the upper and lower quartiles (lighter lines).}
  \label{fig:ion_limit_em}
\end{figure}

Figure~\ref{fig:ion_limit_em} illustrates the consequences of Theorem~\ref{thm:limitingEM} by
plotting the estimates of the parameters $q(1,1)$ and $\mu(1)$ obtained by the batch EM algorithm,
as a function of the number of EM iterations, for two different sample sizes: $n=500$ (dotted
lines) and $n=8000$ iterations. To give an idea of the variability of the estimates,
Figure~\ref{fig:ion_limit_em} feature the median estimate (bold line) as well as the lower and
upper quartiles (lighter curves) for both sample sizes. The first striking observation is the slow
convergence of EM in this case, which requires about 50 iterations or so to reach decent
estimates of the parameters. When comparing the curves corresponding to the two samples sizes, it
is also obvious that while the variability is greatly reduced for $n=8000$ compared to $n=500$, the
median learning curve is very similar in both cases. Furthermore, the plots corresponding to
$n=8000$ provide a very clear picture of the deterministic limiting EM trajectory, whose existence
is guaranteed by Theorem~\ref{thm:limitingEM}.

\begin{figure}[hbtp]
  \centering
  \includegraphics[width=0.6\textwidth]{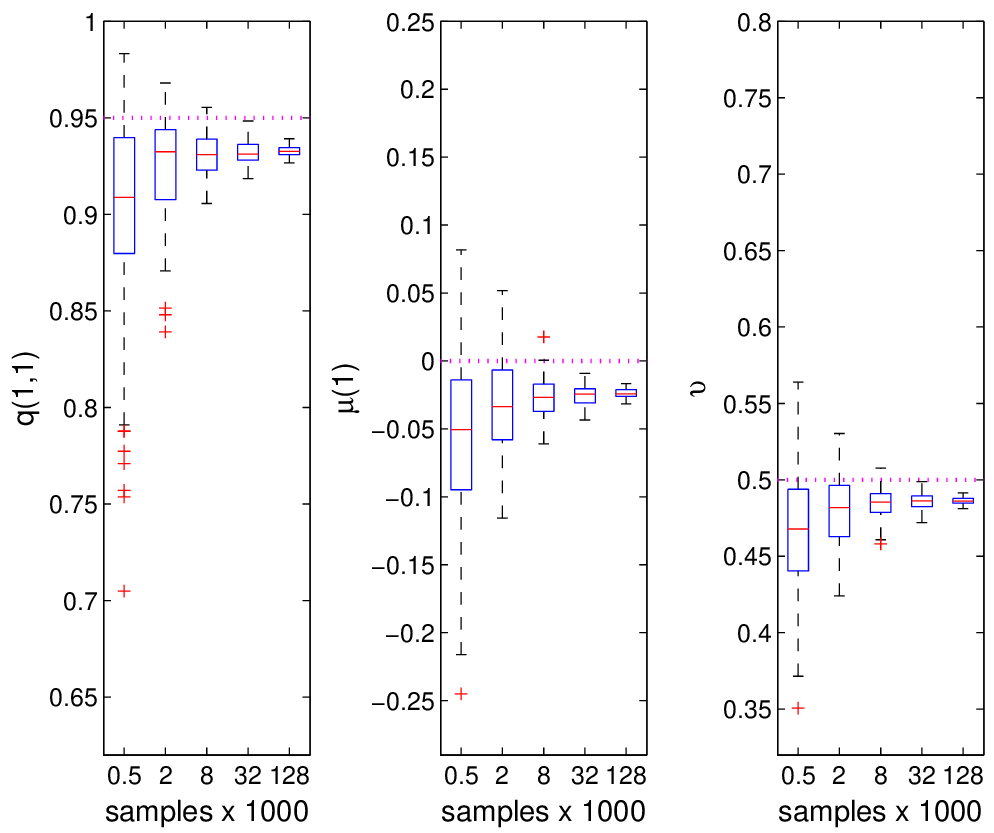}  
  \caption{Estimation results when using 50 batch EM iterations. From left to right, estimated values of $q(1,1)$, $\mu(1)$ and $\upsilon$ for values of $n$ ranging from 0.5 to 128 thousands of samples. Box and whiskers plot based on 100 independent runs.}
  \label{fig:ion_batch}
\end{figure}

\begin{figure}[hbtp]
  \centering
  \includegraphics[width=0.6\textwidth]{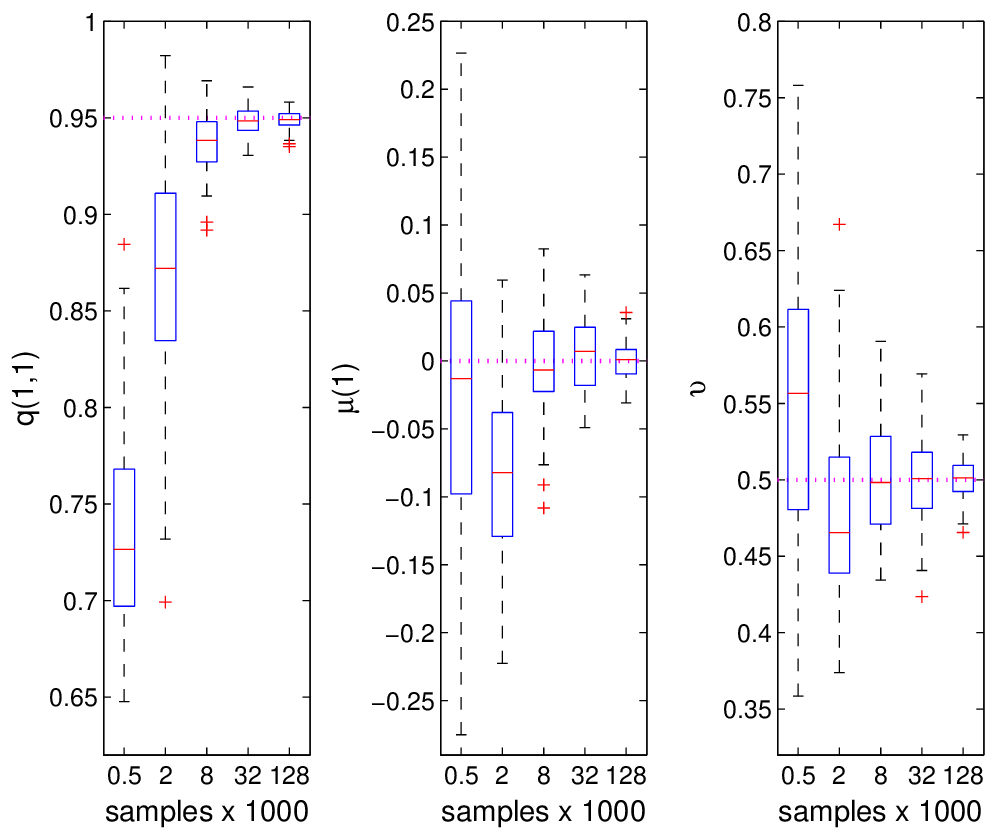}  
  \caption{Estimation results when using the online EM algorithm with $\gamma_n = n^{-0.6}$. From left to right, estimated values of $q(1,1)$, $\mu(1)$ and $\upsilon$ for values of $n$ ranging from 0.5 to 128 thousands of samples. Box and whiskers plot based on 100 independent runs.}
  \label{fig:ion_comp1}
\end{figure}

Indeed, the large sample behavior of the batch EM algorithm is rather disappointing as using a
fixed number of iteration of EM does involve a computational cost that grows proportionally to $n$
but will converge, as $n$ grows, to a deterministic limit which only depends on the parameter
initialization. This is all the more regrettable that from a statistical perspective, it is expected
that the true maximum likelihood estimator converges, at rate $n^{-1/2}$ towards the actual
value $\theta_\star$ of the parameter. This behavior of the batch EM algorithm is illustrated on
Figure~\ref{fig:ion_batch} which displays, from left to right, the estimation results for the
parameters associated with the first component ($q(1,1)$, $\mu(1)$), as in
Figure~\ref{fig:ion_limit_em}, together with the noise variance $\upsilon$ (rightmost display). The results obtained on the 100 independent
runs are here summarized as box and whiskers plots. Figure~\ref{fig:ion_batch}, which
should be compared with Figure~\ref{fig:ion_comp1} below, shows that when using a potentially large
(here, 50) but fixed number of iterations the variability of the batch EM estimates does decrease
but the accuracy does not improve as $n$ grows. Clearly, statistical consistency could only be
achieved by using more batch EM iterations as the number $n$ of observations grows.

\begin{figure}[hbtp]
   \centering
   \includegraphics[width=0.6\textwidth]{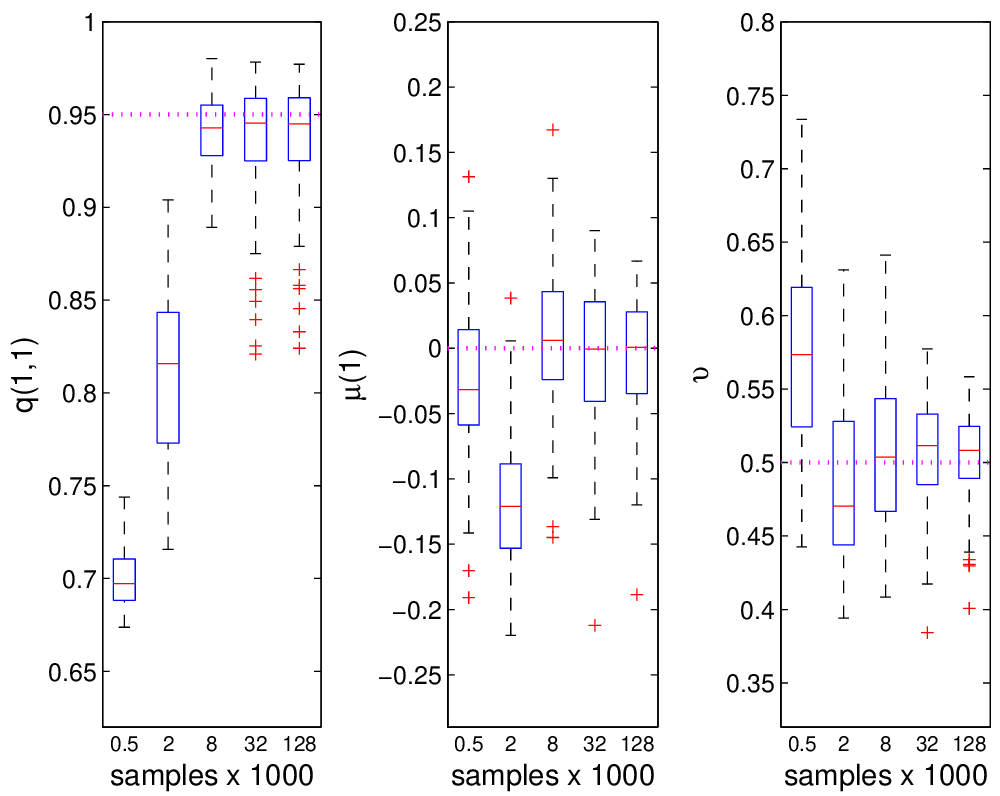}  
   \caption{Estimation results when using the online EM algorithm with $\gamma_n = 0.01$ for $n \leq n_0$ and $\gamma_n = 0.5(n-n_0)^{-1}$ for values of $n > n_0$, with $n_0 = 10000$. From left to right, estimated values of $q(1,1)$, $\mu(1)$ and $\upsilon$ for values of $n$ ranging from 0.5 to 128 thousands of samples. Box and whiskers plot based on 100 independent runs.}
   \label{fig:ion_comp1_1m}
\end{figure}

In contrast, Figure~\ref{fig:ion_comp1} which corresponds to the online EM algorithm
(Algorithm~\ref{alg:MC+noise}) used with $\gamma_n = n^{-0.6}$ does suggest that online EM
estimation is consistent. For the smallest sample sizes ($n= 500$ or $n= 2000$), the estimation
variance is still quite large and the online estimates are not as good as those obtained using 50
batch EM iterations. But for sample sizes of $n=8000$ and larger, the online EM estimates are
preferable despite their somewhat larger variance. In this application, the choice of a slowly
decreasing step-size appears to be of utmost importance. In particular, the choice $\gamma_n =
n^{-1}$, despite its strong analogy with the batch EM case (see Section~\ref{sec:alg:HMMs} as well
as \citealp{neal:hinton:1999}) provides estimates that are not robust enough with respect to the
choice of the initial parameter guess. Simple step-size schemes of the form $\gamma_n =
n^{-\alpha}$ perform very poorly in this example when $\alpha$ is set to values higher than
0.8. Figure~\ref{fig:ion_comp1_1m} features a more complex choice of step-sizes which was
hand-tuned based on pilot runs. Even with this preliminary phase of step-size tuning ---which would
hardly be feasible in real-life applications of the method--- it is observed on
Figure~\ref{fig:ion_comp1_1m} that the results are not significantly improved compared to the
simple choice of $\gamma_n = n^{-0.6}$ as in Figure~\ref{fig:ion_comp1}. The comparatively higher
variability observed for the largest sample sizes (32 and 128 thousands) on
Figure~\ref{fig:ion_comp1_1m} is caused by a very slow reduction of the bias of the parameter
estimates when $\gamma_n=n^{-1}$. In contrast, parameter trajectories corresponding to the choice of $\gamma_n =
n^{-0.6}$ look much less smooth but provide more reliable estimates, without requiring
model-specific tuning of the step-sizes. This observation ---which is also true, but to a lesser
extent, in the case of independent observations--- is certainly a consequence of the temporal
dependence between the observations and, correlatively, of the time needed for the filtering and
smoothing relations to forget their initial state.

\begin{table}[hbt]
  \centering
  \begin{tabular}{|l|c|c|} \hline
    Method & MATLAB 7.7 & OCTAVE 3.0 \\ \hline
    Online EM & 1.57 & 5.66 \\
    Batch EM (one iteration, recursive) & 1.24 & 3.98 \\
    Batch EM (one iteration, forward-backward) & 0.31 & 2.94 \\ \hline
  \end{tabular}
  \caption{Computing times in seconds for a record of length $n=10000$ observations (Intel Core 2 E6600 2.4 GHz processor).}
  \label{tab:cpu}
\end{table}

The comparison between Figures~\ref{fig:ion_batch} and~\ref{fig:ion_comp1} is not
meant to be fair in terms of computing time, as shown by Table~\ref{tab:cpu}\footnote{The
  MATLAB/OCTAVE code used for the simulations is very simple and is available as supplementary
  material. It is mostly vectorized, except for a loop through the observations, and hence it is
  expected that the differences in running times are indeed representative, despite the use of an
  interpreted programming language.}. The 50 batch EM iterations used to produce
Figure~\ref{fig:ion_batch} take about 10 to 40, depending on the implementation of batch EM, times
longer than for the corresponding online estimates of Figure~\ref{fig:ion_comp1}. Being fair in
this respect would have mean using just five batch EM iterations which, as can be guessed from
Figure~\ref{fig:ion_limit_em}, is not competitive with online EM, even for the smallest sample
sizes. Note that a different option would have been to also consider running the online EM
algorithm several times on the same batch data, following \cite{neal:hinton:1999}. In the case of
hidden Markov models however, this way of using the online EM algorithm for fixed-sample maximum
likelihood estimation of the parameters appears to be less straightforward than in the case of
i.i.d. data and has not been considered.

The two batch EM implementations featured in Table~\ref{tab:cpu} correspond, respectively, to the
use of the recursive form of smoothing based on Proposition~\ref{prop:recursive} and to the usual
forward-backward form of smoothing. The former implementation is obviously related to the online EM
algorithm, which explains that both of them lead to rather similar running times. As discussed in
Section~\ref{sec:discuss}, due to the fact that the whole $m \times m$ transition matrix $q$ is
here used as a parameter, the numerical complexity of the online EM algorithm and of the recursive
implementation of batch EM scale as $m^4$, compared to $m^2$ only for the batch EM algorithm when
implemented with forward-backward smoothing. Hence, it is to be expected that the forward-backward
implementation of batch EM would be even more advisable for models with more than $m=2$
states. On the other hand, when $m$ is large it is usually not reasonable to parameterize the
transition matrix $q$ by its individual entries.

\begin{figure}[hbtp]
  \centering
  \includegraphics[width=0.6\textwidth]{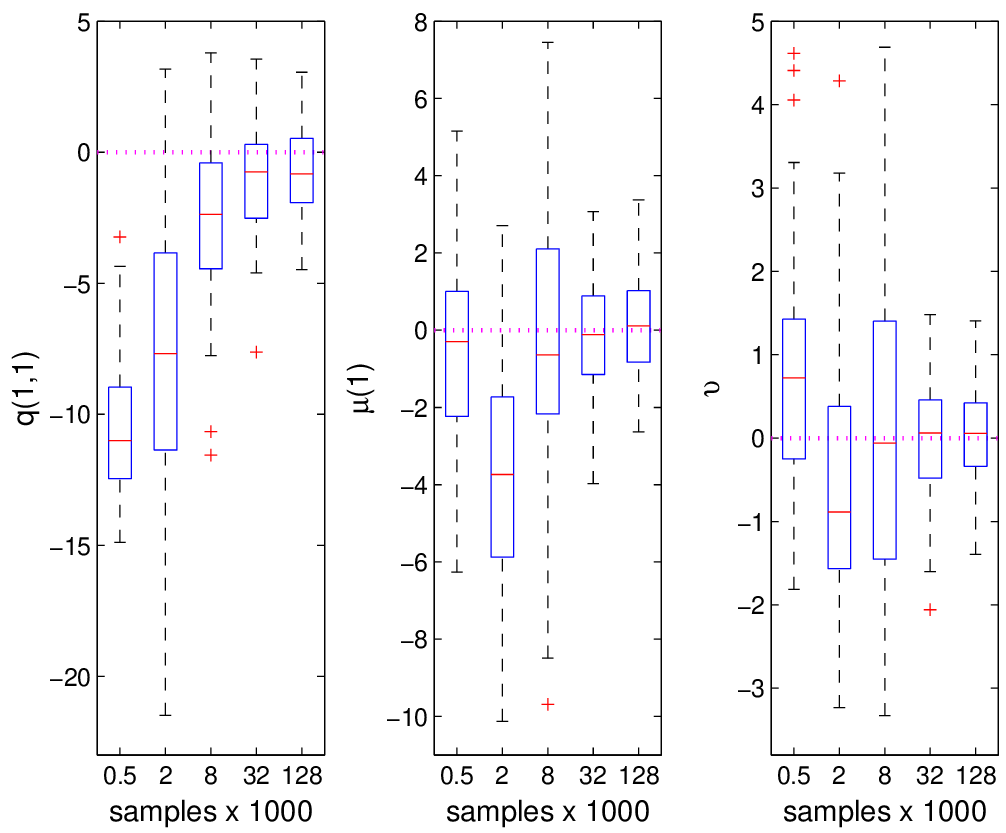}  
  \caption{Estimation results when using the online EM algorithm with $\gamma_n = n^{-0.6}$ with Polyak-Ruppert averaging started after $n=8000$. From left to right, estimated values of $q(1,1)$, $\mu(1)$ and $\upsilon$ for values of $n$ ranging from 0.5 to 128 thousands of samples. The estimated values are centered and scaled so as to be comparable with a unitary asymptotic standard deviation. Box and whiskers plot based on 100 independent runs.}
  \label{fig:ion_comp1_avg}
\end{figure}

In order to provide a more detailed idea of the asymptotic performance of the algorithm,
Figure~\ref{fig:ion_comp1_avg} displays results similar to those of Figure~\ref{fig:ion_comp1} but
centered and scaled as follows. Each parameter estimates, say $\hat{\theta}_n$ is represented as
$\sqrt{n}(\hat{\theta}_n-\theta_\star)$ and further scaled by the asymptotic standard deviation of
$\theta$ deduced from the inverse of the Fisher information matrix. The Fisher information matrix
has been estimated numerically by applying Fisher's identity to~(\ref{eq:Q_EM:MCinNoise}) so as to
obtain
\begin{align*}
  & \frac1n \nabla_{q(i,j)} \log \ell_\theta(Y_{0:n}) = \frac{S_n^q(i,j)}{q(i,j)} - \frac{S_n^q(i,m)}{q(i,m)} \quad \text{(for $1 \leq j < m$)} \eqsp , \\
  & \frac1n \nabla_{\mu(i)} \log \ell_\theta(Y_{0:n}) = \frac{S_{n,1}^g - \mu(i) S_{n,0}^g}{\upsilon} \eqsp , \\
  & \frac1n \nabla_{v} \log \ell_\theta(Y_{0:n}) = \sum_{i=1}^m \frac{1}{2v^2} \left(S^{g}_{n,2}(i;\theta) - 2\mu(i) S^{g}_{n,1}(i) + \mu^2(i)  S^{g}_{n,0}(i) \right) \eqsp .
\end{align*}
The information matrix has then been estimated by averaging the gradient computed in $\theta_\star$ 
for 100 independent sequences of length one million simulated under $\theta_\star$.

Additionally, Figure~\ref{fig:ion_comp1_avg} also displays results that have been post-processed
using Polyak-Ruppert averaging \citep{polyak:1990,ruppert:1988}. In Figure~\ref{fig:ion_comp1_avg}, Polyak-Ruppert
averaging is used starting from $n_{\operatorname{avg}} = 8000$. That is, for $n > 8000$,
$\hat{\theta}_n$ is replaced by $1/(n-8000) \sum_{i=8001}^n \hat{\theta}_n$. For time indices $n$
smaller than 8000, averaging is not performed and the estimates are thus as in
Figure~\ref{fig:ion_comp1}, except for the centering and the scaling. Under relatively mild
assumptions, averaging has been shown to improve the asymptotic rate of convergence of stochastic
approximation algorithm making it possible to recover the optimal rate of convergence of $n^{-1/2}$. At least for $\mu(1)$ and $\upsilon$, Figure~\ref{fig:ion_comp1_avg} suggests that in
this example the proposed algorithm does reach asymptotic efficiency, i.e., becomes asymptotically
equivalent to the maximum likelihood estimator. For $q(1,1)$ the picture is less clear as the
recentered and scaled estimates present a negative bias which disappears quite slowly. This effect
is however typical of the practical trade-off involved in the choice of the index
$n_{\operatorname{avg}}$ where averaging is started. To allow for a significant variance reduction,
$n_{\operatorname{avg}}$ should not be too large. On the other hand, if averaging is started too
early, forgetting of the initial guess of the parameters occurs quite slowly. In the present case,
the negative bias visible on the left panel of Figure~\ref{fig:ion_comp1_avg} is due to
$n_{\operatorname{avg}}$ being too small (see corresponding panel in
Figure~\ref{fig:ion_comp1}). Although, this could be corrected here by setting
$n_{\operatorname{avg}}$ to twenty thousands or more, it is important to underline that optimally
setting $n_{\operatorname{avg}}$ is usually not feasible in practice.

\section{Conclusions}
The algorithm proposed in this paper for online estimation of HMM parameters is based on two
ideas. The first, which is inspired by \cite{sato:2000} and \cite{cappe:moulines:2009}, consists in
reparameterizing the model in the space of sufficient statistics and approximating the limiting EM
recursion by a procedure resembling stochastic approximation. Theorem~\ref{thm:limitingEM} provides a first
argument demonstrating that this idea can also be fruitful in the case of HMMs. The second element is
more specific to HMMs and relies on the recursive implementation of smoothing
computations for sum functionals of the hidden state which is provided by Proposition~\ref{prop:recursive}. As discussed in Section~\ref{sec:alg:HMMs}, this possibility requires
that the auxiliary quantity $\rho_{n,\nu,\theta}$ defined in (\ref{eq:stat:def}) be approximated
during the course of the algorithm.

Although the performance reported in Section~\ref{sec:gausHMM} is encouraging, there are several
questions raised by this approach. The first is of course the theoretical analysis of the
convergence of Algorithm~\ref{alg:generic}, which is still missing. Although originally inspired by
stochastic approximation ideas, it seems that Algorithm~\ref{alg:generic} would be difficult to
analyze using currently available stochastic approximation results due to the backward kernel operator $\hat{r}_{n+1}(x'|x)$
involved in (\ref{eq:stat:update:online}). As discussed in Sections~\ref{sec:discuss}
and~\ref{sec:exper}, the proposed algorithm may become less attractive, from a computational point
of view, when used in models with many distinct state values.
In such cases and, more generally, in cases where the state-space $\mathcal{X}$
of the hidden chain is no longer finite, a promising approach consists in using
some form of Monte Carlo computation to approximate
(\ref{eq:stat:filter:online}) and
(\ref{eq:stat:update:online}). \cite{cappe_09-onlinesmc-ssp} and
\cite{delmoral_doucet_singh_09} report encouraging first results in that
direction.

\appendix

\section{Proofs} 
\label{appendix}

Theorem~\ref{thm:limitingEM} mainly relies on the use of a two-sided forgetting result which
is first proved in Corollary~\ref{cor:2sided} below. This result generalizes the one-sided
forgetting bounds of \cite{douc:moulines:ryden:2004} and allows conditioning with respect to both
past and future observations, which is required for studying the asymptotic behavior
of~\eqref{eq:Fisher-HMM} and related quantities. The proof of Theorem~\ref{thm:limitingEM} then
mostly relies on the results of \cite{douc:moulines:ryden:2004}.

\begin{lemma}
 \label{lem:bayes}
 Given $q$ a transition matrix on the finite set $\mathcal{X}$ such that $q(x,x') \geq \epsilon > 0$ and $\alpha$ and $\beta$ probabilities on $\mathcal{X}$, define
 \[
   J_{\alpha,q,\beta}(x,x') = \frac{\alpha(x) q(x,x') \beta (x')}{\sum_{x,x'\in\mathcal{X}^2}\alpha(x) q(x,x') \beta (x')} \eqsp .
\]
Then
\begin{equation}
  \|J_{\alpha_1,q,\beta_1} - J_{\alpha_2,q,\beta_2}\|_{1} \leq \frac{1}{\epsilon} \left( \| \alpha_1 - \alpha_2\|_{1} + \| \beta_1 - \beta_2\|_{1} \right) \eqsp ,
\label{eq:uioaec}
\end{equation}
where $\|\mu\|_{1} = \sum_{x} |\mu(x)|$ denotes the $L^1$ or total variation norm.
\end{lemma}

\begin{proof} Lemma~\ref{lem:bayes} is obviously related to the application of Bayes' formula. Hence, one may apply Lemma 3.6 of~\cite{kuensch:2001} to obtain $\|J_{\alpha_1,q,\beta_1} - J_{\alpha_2,q,\beta_2}\|_{1} \leq \frac{1}{\epsilon} \| \alpha_1 \otimes \beta_1 - \alpha_2 \otimes \beta_2 \|_{1}$. The r.h.s. of~\eqref{eq:uioaec} is obtained by noting that $|\alpha_1(x)\beta_1(x') - \alpha_2(x)\beta_2(x')| \leq |\alpha_1(x) - \alpha_2(x)|\beta_1(x') + |\beta_1(x') - \beta_2(x')|\alpha_2(x)$. 
\end{proof}

\begin{corollary}
  \label{cor:2sided}
  Under the assumptions of Theorem~\ref{thm:limitingEM}, for any function $f$ such that $0 < f
  < \|f\|_{\infty}$ and probabilities $\mu_1$ and $\mu_2$ on $\mathcal{X}^2$, and any index $1 \leq
  t \leq n$,
  \begin{multline*}
  \bigg| \sum_{x,x'\in\mathcal{X}^2} \E_\theta\left[\left. f(X_{t-1},X_t) \right| Y_{0:n}, X_0 =x, X_n = x'\right] \mu_1(x,x') \\
   - \sum_{x,x'\in\mathcal{X}^2} \E_\theta\left[\left. f(X_{t-1},X_t) \right| Y_{0:n}, X_0 =x, X_n = x\right] \mu_2(x,x') \bigg| \leq \frac{\|f\|_{\infty}}{\epsilon} \left(\rho^{t-1} + \rho^{n-t} \right) \eqsp ,    
  \end{multline*}
where $\rho = (1-\epsilon)$.
\end{corollary}

\begin{proof} First apply Lemma~\ref{lem:bayes} to the familiar forward-backward decomposition
  \begin{align*}
  & \alpha_i(x) = \P_\theta(X_{t-1}=x|Y_{0:t-1}, X_0
  = x_{0,i}) \eqsp , \\
  & \beta_i(x') \propto \P_\theta(Y_{t:n}, X_n = x_{n,i}|X_{t}=x') \eqsp ,    
  \end{align*}
 for $i=1,2$ (where the normalization factor in the second equation is determined by the constraint $\sum_{x\in \mathcal{X}} \beta_i(x) = 1$) to obtain
  \begin{multline*}
  \bigg| \E_\theta\left[\left. f(X_{t-1},X_t) \right| Y_{0:n}, X_0 =x_{0,1}, X_n = x_{n,1}\right] \\
   - \E_\theta\left[\left. f(X_{t-1},X_t) \right| Y_{0:n}, X_0 = x_{0,2}, X_n = x_{n,2}\right] \bigg| \leq \frac{\|f\|_{\infty}}{\epsilon} \left(\|\alpha_1-\alpha_2\|_1 + \|\beta_1-\beta_2\|_1 \right) \eqsp,
  \end{multline*}
observing that $\P_\theta(X_{t-1}=x, X_t = x'|Y_{0:t-1}, X_0 = x_{0,i}, X_n = x_{n,i}) = J_{\alpha_i,q_\theta,\beta_i}$. Next use, the one-sided forgetting bounds of \cite{douc:moulines:ryden:2004} (Corollary 1 and Eq. (20)) to obtain $\|\alpha_1-\alpha_2\|_1 \leq \rho^{t-1}$ and $\|\beta_1-\beta_2\|_1 \leq \rho^{n-t}$. The result of Corollary~\ref{cor:2sided} follow by the general inequality $|\mu_1(g)-\mu_2(g)| \leq \frac{1}{2} \|\mu_1-\mu_2\|_1 \sup_{z_1,z_2\in\mathcal{Z}^2} |g(z_1)-g(z_2)|$.
\end{proof}

Note that the fact that the backward function $\beta_i(x') = \P_\theta(Y_{t:n}, X_n = x_{n,i}|X_{t}=x')$
may be normalized to a pseudo-probability, has been used in the above proof. This is generally not
the case outside of the context where $\mathcal{X}$ is finite
\citep{cappe:moulines:ryden:2005,briers:doucet:maskell:2010} but it is easily checked that
Corollary~\ref{cor:2sided} holds in greater generality under the ``strong mixing conditions''
discussed in Section 4.3 
of~\cite{cappe:moulines:ryden:2005}.

\begin{proof}[Proof of Theorem~\ref{thm:limitingEM}]
  Corollary~\ref{cor:2sided} implies that
  \[
  \left|\E_{\theta}\left[\left. s(X_{-1},X_0,Y_0) \right| Y_{-n:n} \right] -
    \E_{\theta}\left[\left. s(X_{-1},X_0,Y_0) \right| Y_{-m:m} \right]\right| \leq \frac2\epsilon
  \rho^n M(Y_0) ,
  \]
  for $m \geq n$, where $M(y) \geq \sup_{x,x'} |s(x,x',y)|$. $M(y)$ may be chosen as
  $M(y) = \sum_x 1 + |s^g(x,y)|$ due to the fact that $s^q(x,x')$ is a vector of indicator
  functions when $\mathcal{X}$ is finite. As, $\theta_\star$ lies in the interior of $\Theta$,
  standard results on exponential family imply that $M(Y_0)$ has finite first and second order moments
  under $\P_{\theta_\star}$. Hence, the a.s. limit of $\E_{\theta}\left[\left. s(X_{-1},X_0,Y_0)
    \right| Y_{-m:m} \right]$ as $m\to\infty$, which is denoted by
  $\E_{\theta}\left[\left. s(X_{-1},X_0,Y_0) \right| Y_{-\infty:\infty} \right]$, exists and has
  finite expectation under $\P_{\theta_\star}$. Similarly,
  \[
  \frac{1}{n} \sum_{t=1}^n \left(\E_{\nu,\theta}\left[\left. s(X_{t-1},X_t,Y_t) \right| Y_{0:n}
    \right] - \E_{\theta}\left[\left. s(X_{t-1},X_t,Y_t) \right| Y_{-\infty:\infty} \right] \right)
  \leq \frac{1}{n\epsilon} \sum_{t=1}^n \left( \rho^{t-1} + \rho^{n-t} \right) M(Y_t) \eqsp .
  \]
  As $\E_{\theta_\star}[M(Y_0)^2] < \infty$, standard applications of Markov inequality
  and Borel-Cantelli Lemma imply that the r.h.s. of the above expression tends $\P_{\theta_\star}$--a.s. to zero. Hence,
  the quantities $\frac1n \E_{\nu,\theta}\left[\sum_{t=1}^n \left. s(X_{t-1},X_t,Y_t) \right|
    Y_{0:n} \right]$ and $\frac1n \sum_{t=1}^n \E_{\theta}\left[\sum_{t=1}^n
    \left. s(X_{-1},X_0,Y_0) \right| Y_{-\infty:\infty} \right]$ have the same limit, where the
  latter expression converges to $\E_{\theta_\star}\left(\E_{\theta}\left[\left. s(X_{-1},X_0,Y_0)
      \right| Y_{-\infty:\infty} \right]\right)$ by the ergodic theorem. This proves the first
  assertion of Theorem~\ref{thm:limitingEM}.
  
  For the second statement, one can check that the assumptions of Theorem~\ref{thm:limitingEM}
  imply (A1)--(A3) of \cite{douc:moulines:ryden:2004} as well as a form of
  (A6)--(A8)\footnote{Theorem 3 of \cite{douc:moulines:ryden:2004} deals with the Hessian of the
    normalized log-likelihood. As we are only concerned with the gradient here, one can drop the
    second order conditions in (A6)--(A7). Furthermore, as the assumption of
    Theorem~\ref{thm:limitingEM} are supposed to hold uniformly on $\Theta$, the set $G$ can be
    dropped in (A6)--(A7), which provides a law of large number for the score that holds for all
    values of $\theta \in \Theta$.}. Hence, proceeding as in proof of Theorem 3 of
    \cite{douc:moulines:ryden:2004}, shows that~(\ref{eq:Fisher-HMM}) converge a.s. to the
  gradient $\nabla_\theta c_{\theta_\star}(\theta)$ of the limiting contrast defined
  in~(\ref{eq:HMM:contrast}). Eq.~\eqref{eq:gradient_expform} combined with the previous result
  then shows that parameter values $\theta$ for which $\nabla_\theta c_{\theta_\star}(\theta)$
  vanishes are also such that  $\nabla_\theta\psi(\theta) \left\{
    \E_{\theta_\star}\left(\E_{\theta}\left[\left. s(X_{-1},X_0,Y_0) \right| Y_{-\infty:\infty}
      \right]\right) \right\} -\nabla_\theta A(\theta) = 0$, that is,
  $\bar{\theta}\left\{\E_{\theta_\star}\left(\E_{\theta}\left[\left. s(X_{-1},X_0,Y_0) \right|
        Y_{-\infty:\infty} \right]\right)\right\} = \theta$.
\end{proof}

In order to prove Corollary~\ref{corr:rho_freeze}, one first needs the following simple lemma.

\begin{lemma}
  \label{eq:lemma:wghtsum}
  Let $(e_k)_{k\geq 0}$ denote a deterministic sequence and $(\gamma_k)_{k\geq 1}$ a sequence of
  step-sizes such that $\gamma_k \in [0,1]$, $\sum_{k\geq 1} \gamma_k = \infty$, and, $\sum_{k\geq 1}
  \gamma_k^2 < \infty$; define $r_{k} = (1-\gamma_{k}) r_{k-1} + \gamma_{k} e_k$ for $k \geq
  1$. Then
  \begin{enumerate}
    \item [(i)] $r_n = \sum_{k=1}^n \omega_k^n e_k + \omega_0^n r_0$, where $\omega_k^n = \gamma_k \prod_{j=1}^n (1-\gamma_j)$.
    \item [(ii)] $r_n$ and $\frac1n \sum_{k=1}^n e_k$ have the same limit.
   \end{enumerate}
\end{lemma}

Lemma~\ref{eq:lemma:wghtsum}-(ii) can be proved using Abel transformation noticing that
$\sum_{k=1}^n \omega_k^n$ is by definition equal to one and that for the range of step-sizes
considered here, $(\omega_k^n)_{1\leq k\leq n}$ is a non-decreasing sequence (it is equal to
$n^{-1}$ when $\gamma_k = k^{-1}$ and strictly increasing for sequences of step-sizes that decrease
more slowly). Lemma~\ref{eq:lemma:wghtsum} shows that we may concentrate on the case where
$\gamma_k = k^{-1}$ to prove Corollary~\ref{corr:rho_freeze}. Applying
Proposition~\ref{prop:recursive}, $\hat{\rho}_n$ is then equal under parameter freeze to
$\rho_{n,\nu,\theta}$ defined in~(\ref{eq:stat:def}). In light of Corollary~\ref{cor:2sided},
$\rho_{n,\nu,\theta}$ has the same behavior as the term $\frac{1}{n} \sum_{t=1}^n
\E_{\nu,\theta}\left[\left. s(X_{t-1},X_t,Y_t) \right| Y_{0:n} \right]$ that was analyzed above
when proving Theorem~\ref{thm:limitingEM}.

\bibliography{article} 

\end{document}